\documentstyle[12pt,aaspp4,amsmath,epsfig]{article} 

\lefthead{Salmonson and Wilson}
\righthead{Neutrino Heating}

\begin{document}

\title{ General Relativistic Augmentation of Neutrino Pair
Annihilation Energy Deposition Near Neutron Stars }

\author{Jay D. Salmonson\altaffilmark{1,2} and James R. Wilson}
\affil{Lawrence Livermore National Laboratory, Livermore, CA 94550}

\altaffiltext{1}{Department of Applied Science, University of California, Davis, CA 95616}
\altaffiltext{2}{e-mail: salmonson@llnl.gov}

\begin{abstract}

General relativistic calculations are made of neutrino-antineutrino
annihilation into electron-positron pairs near the surface of a
neutron star. It is found that the efficiency of this process is
enhanced over the Newtonian values up to a factor of more than 4 in
the regime applicable to Type II supernovae and by up to a factor of
30 for collapsing neutron stars.

\end{abstract}

\keywords{ neutrinos---radiative transfer---stars:supernovae }

\section{Introduction}

It has long been realized that the reaction $\nu + {\overline{\nu}}
\rightarrow e^+ + e^-$, near the surface of a hot neutron star, is of
considerable importance to type II supernova dynamics.  At late times,
in particular, several percent of the neutrino luminosity ($L_{\nu}
\sim 10^{52}$ ergs/sec) from the hot protoneutron star is deposited
into the stellar envelope via $\nu{\overline{\nu}} \rightarrow e^+e^-$
and neutrino-lepton scattering.  This energy deposition, together with
neutrino-baryon capture, significantly augments the neutrino heating
of the envelope in a successful supernova via the so-called ``delayed
shock mechanism'' (\cite{jay:bw85,jay:bethe90}).  The late time
heating of the envelope is most important for the r-process
nucleosynthesis.

Another case of interest stems from neutron star collapse.  This could
occur either from accretion (\cite{jay:bvdh}) directly from a
companion or from plunging into the interior of a companion
(\cite{jay:fbh96}).  A third possibility is suggested in recent
studies of close neutron star binaries near their last stable orbit
(\cite{jay:wmm96,jay:mw97}). This work indicates that neutron stars in
such systems can compress, heat and emit a neutrino luminosity as high
as $10^{53}$ ergs/second.  A fraction of these neutrinos should be
converted to $e^+e^-$ pairs via $\nu{\overline{\nu}} \rightarrow
e^+e^-$.  In all of these scenarios, understanding the efficiency of
this reaction is crucial to modeling the energy dynamics and
observable emission, such as the formation of gamma-ray bursts.

Previous calculations of reaction efficiencies near neutron stars have
been based upon Newtonian gravity (Goodman, Dar \& Nussinov 1987,
Cooperstein, van den Horn \& Baron 1986); i.e. they assume $\frac{2 G
M}{c^2 R} \ll 1$ where $M$ is the gravitational mass of the neutron
star and $R$ is the distance scale.  In fact $\frac{2 G M}{c^2 R}$ can
be $\sim 0.4$ for supernova calculations and $\sim 0.7$ for collapsing
neutron stars (\cite{jay:mw97}) and the effects of gravity cannot be
ignored.

In this paper we develop a relativistic analytical model to compute
the efficiency of $\nu{\overline{\nu}} \rightarrow e^+e^-$ in a
high-gravity environment such as near a neutron star.  We find that
relativistic enhancements to the Newtonian reaction rate can be large;
up to factors of 10.

\section{ Neutrino Annihilation }

Newtonian calculations of the $\nu{\overline{\nu}} \rightarrow e^+e^-$
energy deposition rate near a hot neutron star were done by
\cite{jay:cvb} and \cite{jay:gdn}.  The energy deposition per unit
time, per volume (Goodman {\it et al.} 1987) is
\begin{equation}
\dot{q}(r) = \iint f_\nu({\bf p}_\nu,r) 
f_{\overline{\nu}}({\bf p}_{\overline{\nu}},r)
 \{\sigma |{\bf v}_\nu - {\bf v}_{\overline{\nu}} | \varepsilon_\nu 
\varepsilon_{\overline{\nu}} \} { \varepsilon_\nu + \varepsilon_{\overline{\nu}}
\over \varepsilon_\nu \varepsilon_{\overline{\nu}} } d^3{\bf p}_\nu 
d^3{\bf p}_{\overline{\nu}} 
\label{E:phasesum}
\end{equation}

\noindent
where $f_\nu$ and $f_{\overline{\nu}}$ are number densities in phase
space, ${\bf v}_\nu$ is the neutrino velocity, and $\sigma$ is the
rest frame cross section.  The term in curly brackets in equation
(\ref{E:phasesum}) is Lorentz invariant.  Thus it can be calculated in
the center-of-mass frame to be
\begin{equation} 
\{\sigma |{\bf v}_\nu - {\bf v}_{\overline{\nu}} | \varepsilon_\nu 
\varepsilon_{\overline{\nu}} \} = {D G^2_F c \over 3\pi} (\varepsilon_\nu 
\varepsilon_{\overline{\nu}} - {\bf p}_\nu \cdot 
{\bf p}_{\overline{\nu}} c^2 )^2 ~,
\end{equation}
with $G^2_F = 5.29 \times 10^{-44}$ cm$^2$ MeV$^{-2}$ and
\begin{equation}
D = 1 \pm 4 \sin^2\theta_W + 8 \sin^4 \theta_W ~.
\label{E:D}
\end{equation}

\noindent
Here $\sin^2\theta_W = 0.23$ and the $+$ sign is for
$\nu_e{\overline{\nu}}_e$ pairs, while $-$ is for
$\nu_\mu\overline{\nu}_\mu$ and $\nu_\tau{\overline{\nu}}_\tau$ pairs.
Note that we have assumed that the mass of the electrons is negligible
since, for the applications of interest, the energy of the neutrinos
is greater than 10 MeV.  Now we replace ${\bf p}_\nu c =
\varepsilon_\nu {\boldsymbol \Omega}_\nu$ and $d^3{\bf p}_\nu =
\frac{\varepsilon^2_\nu}{c^3} d\varepsilon_\nu d\Omega_\nu$ where
${\boldsymbol \Omega}_\nu$ is the unit direction vector and
$d\Omega_\nu$ is a solid angle
\begin{equation}
\dot{q} = {D G^2_F \over 3\pi c^5} \Theta(r) \iint f_\nu 
f_{\overline{\nu}} (\varepsilon_\nu + 
\varepsilon_{\overline{\nu}}) \varepsilon_\nu^3 d\varepsilon_\nu
\varepsilon_{\overline{\nu}}^3 d\varepsilon_{\overline{\nu}} ~,
\end{equation}

\noindent
where the angular integrations are represented by
\begin{equation}
\Theta(r) \equiv \iint (1 - {\boldsymbol \Omega}_\nu \cdot 
{\boldsymbol \Omega}_{\overline \nu})^2 d\Omega_\nu d\Omega_{\overline \nu} ~.
\label{E:Theta}
\end{equation}

\noindent
Thus the energy and angular dependancies have been separated.  The
energy integrals can then be performed in their local frame.  Taking
$f_\nu = {2 \over h^3} (e^{\varepsilon_\nu/kT} + 1)^{-1}$ for fermions
\begin{eqnarray}
\int_0^\infty f_\nu \varepsilon_\nu^3 d\varepsilon_\nu &=& 
{2(kT)^4 \over h^3} {7\pi^4 \over 120} \\
\int_0^\infty f_\nu \varepsilon_\nu^4 d\varepsilon_\nu &=&
{2(kT)^5 \over h^3} {45 \zeta(5) \over 2 }
\end{eqnarray}

\noindent
we finally get
\begin{equation}
\dot{q} = {7DG^2_F\pi^3\zeta(5) \over 2 c^5 h^6} (kT)^9 \Theta(r) 
\propto T(r)^9 \Theta(r)
\label{E:final1}
\end{equation}

\noindent
where $T(r)$ is the temperature measured by a local observer.  All
calculations carried out thus far are in the local frame. In order to
complete the evaluation of $T(r)$ and $\Theta(r)$, gravitational
redshift and path bending must be included.

It is also of interest to calculate the radial momentum density
$\dot{p}$ imparted to the $e^+e^-$ plasma from the $\nu
\overline{\nu}$ luminosity.  This is related to the energy deposition
of Equation (\ref{E:final1}) by

\begin{equation}
\dot{p} = {\dot{q} \over c} {\Phi_p(r) \over \Theta(r)}
\label{E:momentum1} 
\end{equation}

\noindent
where

\begin{equation}
\Phi_p(r) \equiv {1 \over 2} \iint (1 - {\boldsymbol \Omega}_\nu \cdot
{\boldsymbol \Omega}_{\overline \nu})^2 ({\bf \hat{r}} \cdot
({\boldsymbol \Omega}_\nu + {\boldsymbol \Omega}_{\overline \nu}))
d\Omega_\nu d\Omega_{\overline \nu} 
\label{E:phip}
\end{equation}

\noindent
and ${\bf \hat{r}}$ is the unit vector normal to the stellar
surface.

\section{ Bending of Null Geodesics }

From \cite{jay:mtw}, the path of a zero mass particle, a null
geodesic, in Schwarzchild coordinates is given by
\begin{equation}
\biggl({1\over r^2} {dr \over d\phi}\biggr)^2 + {1 - 2M/r \over r^2} =
{1 \over b^2} ~,
\label{E:geodesic}
\end{equation}

\noindent
where $r$ is the distance from the origin, $\phi$ is the longitude,
and $b$ is the impact parameter.  Here the neutron star mass $M$ is
expressed in units of distance; i.e.  ${G \over c^2} = 1$.  All other
units are in cgs.  We wish to express Equation (\ref{E:geodesic}) in
terms of the the angle $\theta$ between the particle trajectory and
the tangent vector to a circular orbit.  In terms of local radial and
longitudinal velocities $v_{\hat{r}}$ and $v_{\hat{\phi}}$ this
becomes
\begin{equation}
\tan\theta = {v_{\hat{r}} \over v_{\hat{\phi}}} = {\sqrt{|g_{rr}|} 
{dr \over d\lambda} \over \sqrt{|g_{\phi\phi}|} {d\phi \over d\lambda}}
= {1 \over r \sqrt{1 - { 2M \over r}}} {dr \over d\phi} ~.
\end{equation}

\noindent
This expression can be substituted into Equation (\ref{E:geodesic})
and simplified to give
\begin{equation}
b = { r \cos \theta_r \over \sqrt{1 - {2M \over r}}} ~.
\end{equation}

\noindent
All points along a single orbit will share the same impact parameter.
Thus a particle emanating from the photosphere at $R$ with a
trajectory denoted by $\theta_R$ will have a trajectory $\theta_r$ at
radius $r$ given by
\begin{equation}
\cos \theta_r = {R \over r} \sqrt{{1 - {2M \over r} \over 1 - {2M \over R}}} 
\cos \theta_R ~.
\label{E:cos}
\end{equation}

\noindent
It is interesting to note that this equation implies a minimum
photosphere radius, i.e. $R_{min} = 3M$, below which a massless
particle (neutrino) emitted tangent to the stellar surface ($\theta_R
= 0$) is gravitationally bound.  The present discussion is restricted
to the domain $R > 3M$.

We may now evaluate Equation (\ref{E:Theta}).  Defining $\mu = \sin
\theta$, then ${\boldsymbol \Omega} = (\mu, \sqrt{1 - \mu^2} \cos
\phi, \sqrt{1 - \mu^2} \sin \phi)$ and $d\Omega = \cos \theta d\theta
d\phi$ we get

\begin{equation}
\begin{split}
\Theta(r) = 4 \pi^2 \int_x^1 \int_x^1 
[1 - 2 \mu_\nu \mu_{\overline \nu} 
+ \mu^2_\nu \mu^2_{\overline \nu} + 
{1 \over 2} (1 - \mu^2_\nu) (1 - \mu^2_{\overline \nu})]\  d\mu_\nu\
d\mu_{\overline \nu} ~,
\end{split}
\end{equation}

\noindent
where we define

\begin{equation}
x \equiv \sqrt{ 1 - \biggl(\frac{R}{r}\biggr)^2 
{1 - {2M \over r} \over 1 - {2M \over R}} } ~.
\label{E:defx}
\end{equation}

\noindent
The result is

\begin{equation}
\Theta(r) = {2 \pi^2 \over 3} (1 - x)^4 (x^2 + 4 x + 5) ~.
\label{E:Thetaresult}
\end{equation}

\noindent
In the Newtonian limit $(M \rightarrow 0)$ this function has the same
form as that in \cite{jay:gdn} and \cite{jay:cvb2}.  In the limit $r
\gg R$ this becomes $\Theta(r) \approx {\pi^2 \over 2}({R \over
r})^8$, so the efficiency of this process falls off strongly with
radius as the streaming particle trajectories become more parallel.

For the momentum density deposition, the integration of Equation
(\ref{E:phip}) is similarly performed and the result is
\begin{equation}
\Phi_p(r) = {\pi^2 \over 6} (1 - x)^4 (8 + 17 x + 12 x^2 + 3 x^3) ~.
\end{equation}
Thus at the surface of the star ($r = R$) we have ${\Phi_p(r) \over
\Theta(r)} = {2 \over 5}$ and the deposited $e^+e^-$ plasma moves with
a high momentum.  At large distances, as $r \rightarrow \infty$, then
${\Phi_p(r) \over \Theta(r)} \rightarrow 1$, as expected.

\subsection{ Gravitational Redshift }

The temperature $T(r)$ in Equation (\ref{E:final1}) is the neutrino
temperature at radius $r$.  This must be put in terms of an observable
quantity such as the observed luminosity $L_\infty$.  To do this we
start by expressing the temperature of the free streaming neutrinos at
radius $r$ in terms of their temperature at the neutrinosphere radius
$R$ using the appropriate graviational redshift.  Temperature, like
energy, varies linearly with redshift

\begin{equation}
T(r) = \sqrt{1 - {2M \over R} \over 1 - {2M \over r}}\ T(R) ~.
\end{equation}

\noindent
The luminosity varies quadratically with redshift

\begin{equation}
L_\infty = \Bigl (1 - {2M \over R}\Bigr) L(R) ~,
\end{equation}

\noindent
and at the neutrinosphere, the local neutrino luminosity for a single
neutrino species is

\begin{equation}
L(R) = L_\nu + L_{\overline{\nu}} = 4 \pi R^2\ {7 \over 4} 
{ac \over 4} T(R)^4 ~.
\end{equation}

\noindent
Combining these, Equation (\ref{E:final1}) becomes

\begin{equation}
\begin{split}
\dot{q} = {7DG^2_F\pi^3\zeta(5) \over 2 c^5 h^6} k^9 
(\case{7}{8}\pi a c)^{-9/4}\ L_\infty^{9/4}\\
\times\  \Theta(r) \Biggr({\sqrt{1 - {2M \over R}} 
\over 1 - {2M \over r}} \Biggl)^{9/2} R^{-9/4} ~.
\label{E:final2}
\end{split}
\end{equation}

\noindent
Equation (\ref{E:final2}) describes the idealized energy deposition in
$e^+e^-$ pairs from the reaction $\nu + \overline{\nu} \rightarrow e^+
+ e^-$ at radius $r$ above a neutron star of radius $R$ and with a
neutrino luminosity $L_\infty$.  The momentum density deposition is
still described by Equation (\ref{E:momentum1}).

\section{ Results }


In order to quantify the total $e^+e^-$ pair energy deposition, we
define $\dot{Q}$ as the integral of $\dot{q}$, given by Equation
(\ref{E:final2}), over proper volume.  $\dot{Q}$ is not an observable
quantity and not a Lorentz invariant in general relativity, however it
is a measure of the total amount of energy converted from neutrinos to
$e^+e^-$ pairs at all radii.  With this quantity we wish to measure
the total amount of local energy deposited via $\nu\overline{\nu}
\rightarrow e^+e^-$ and thus available for neutrino-electron
scattering or other such energy exchange processes.  For this quantity
we get

\begin{equation}
\begin{split}
\dot{Q} &= \int_R^\infty \dot{q} {4\pi r^2 dr \over \sqrt{1 - {2M \over r}}}\\
&= {28 G^2_F \pi^6\zeta(5) \over 3 c^5 h^6} \Biggl({k \over
\sqrt[4]{\case{7}{4} \pi ac}}\Biggr)^9\ D L_\infty^{9/4} 
\Biggl({(1 - {2M \over R})^{3/2} \over R} \Biggr)^{3/2} \\
&\times \int_1^\infty (x - 1)^4 (x^2 + 4 x + 5) 
{y^2 dy \over (1 - {2M \over yR})^5} ~,
\label{E:Qdot}
\end{split}
\end{equation}

\noindent
where $x$ is defined by Equation (\ref{E:defx}) and $y \equiv {r \over
R}$.  This becomes

\begin{equation}
\dot{Q}_{51} = 1.09 \times 10^{-5}\ {\mathcal{F}}\Bigl({M \over R }\Bigr)\ 
D L^{9/4}_{51} R^{-3/2}_6 ~,
\label{E:Qdot51}
\end{equation}

\noindent
where $\dot{Q}_{51}$ and $L_{51}$ are in units of $10^{51}$ ergs/sec,
$R_6$ is the radius in units of 10 km, and we have defined

\begin{equation}
{\mathcal{F}}\Bigl({M \over R }\Bigr) \equiv 3 \Bigl 
(1 - {2M \over R}\biggr)^{9/4} \int_1^\infty (x - 1)^4 (x^2 + 4 x + 5)
{y^2 dy \over (1 - {2M \over yR})^5}.
\end{equation}

\noindent
For the Newtonian case, ${\mathcal{F}}(0) = 1$, and we recover the
result found in \cite{jay:cvb}.  Note that Equation (\ref{E:Qdot51})
is for only a single neutrino flavor.  Hence, Equation (\ref{E:D})
must be evaluated and $L_{51}$ must be summed over each neutrino
flavor in order to get the total energy deposition due to annihilation
of all neutrino flavors.

A key purpose of this paper is to quantify the general relativistic
augmentation of local $e^+e^-$ pair energy deposition compared to the
Newtonian calculations.  In Figure (\ref{intgrlratio}) the ratio
$\dot{Q}_{GR}/\dot{Q}_{Newt} = {\mathcal{F}}({M \over R})$ is plotted
over a range of stellar radii $R/M$.  It is somewhat surprising to
find that the $\nu\overline{\nu}$ annihilation is enhanced by almost a
factor of 2 at the rather large radius $R = 10M$, increasing to a
factor of 4 at $R = 5M$.  As will be discussed, this is the relevant
range for type II supernova models.  In the more extreme relativistic
regime, down to $R = 3M$, pertinent to collapsing neutron stars, this
enhancement rises to nearly a factor of 30.

\begin{center}
\epsfig{file = 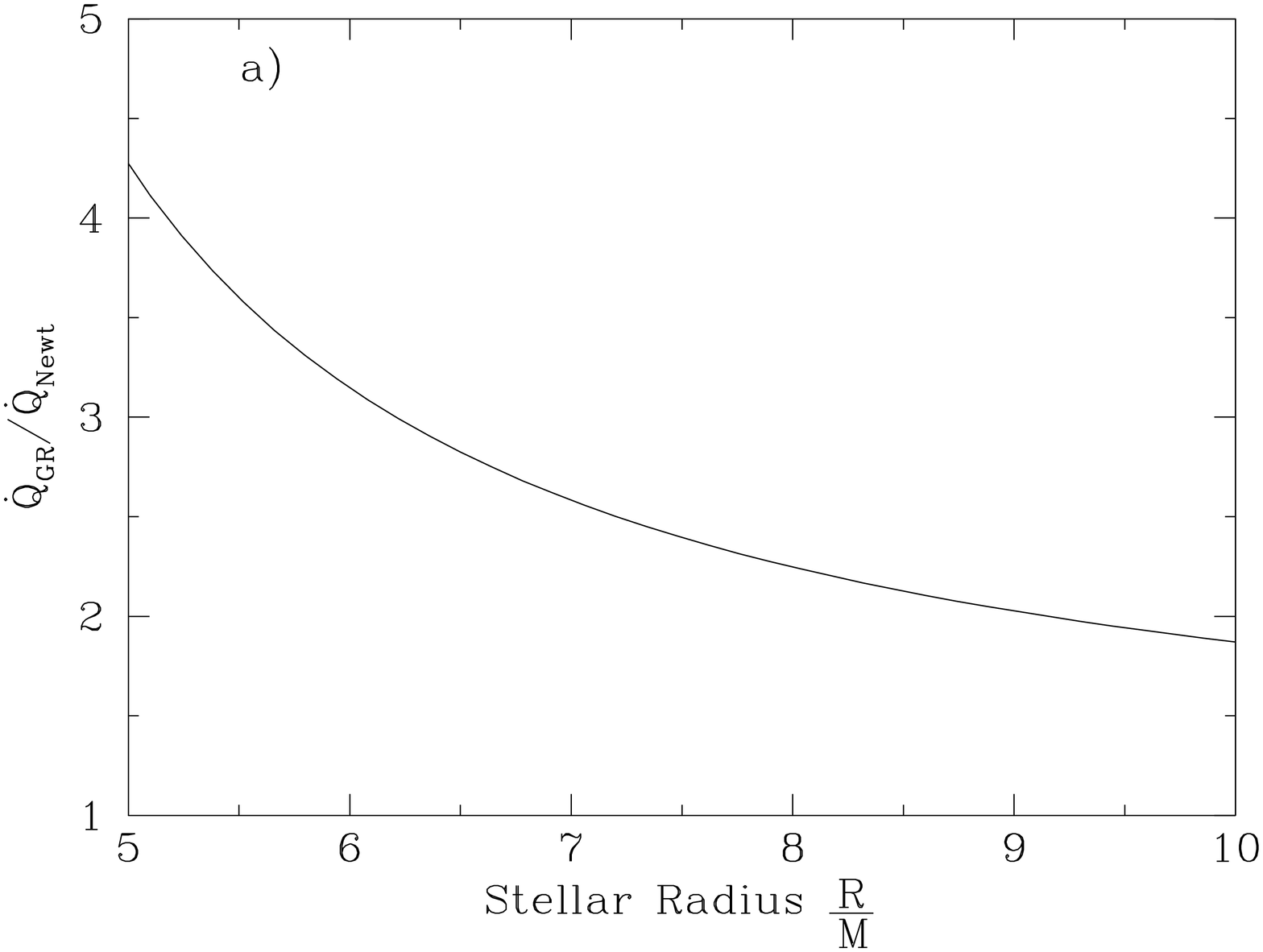, height = 15 cm, width = 7.5 cm, angle = -90}
\epsfig{file = 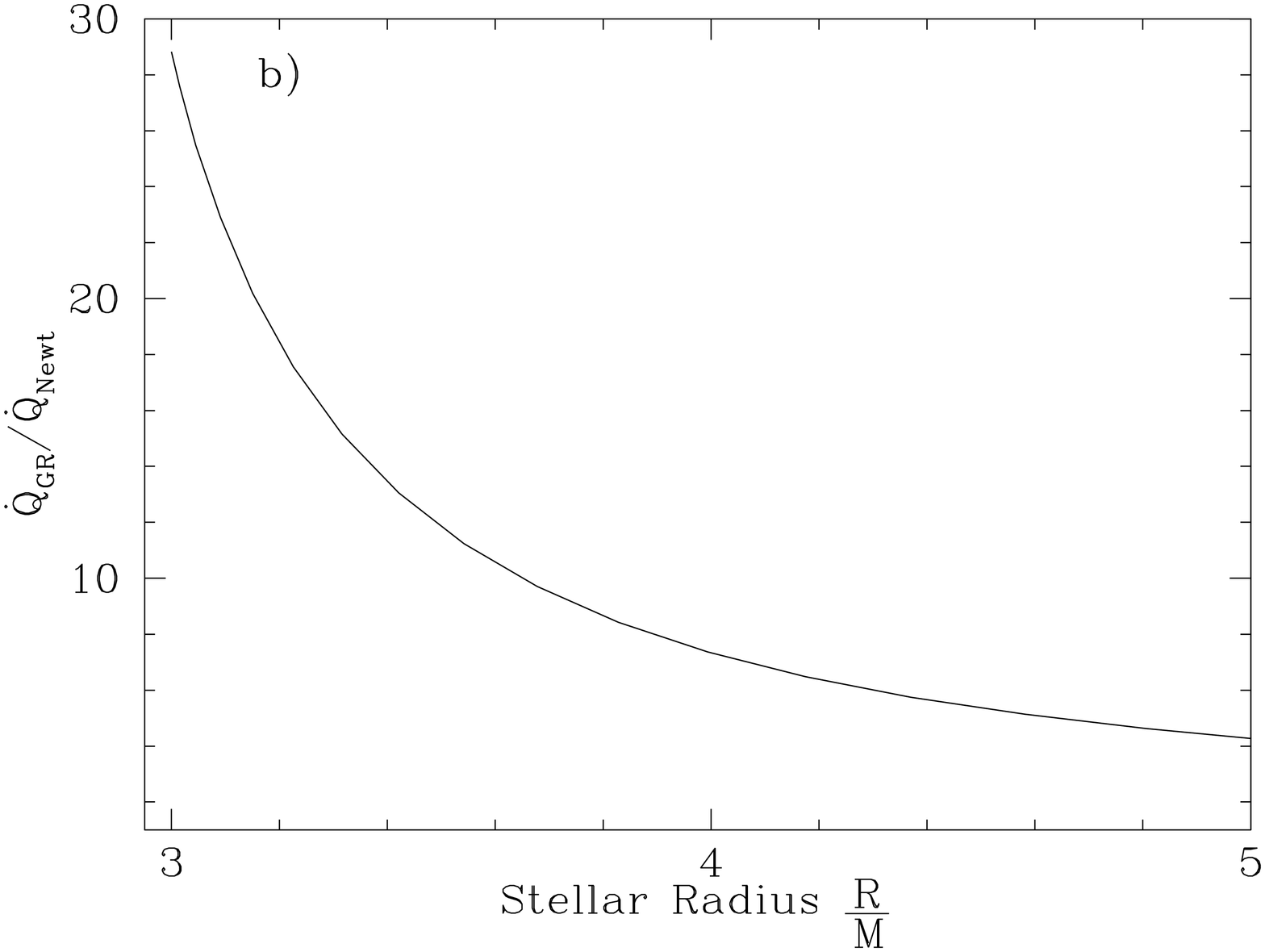, height = 15 cm, width = 7.5 cm, angle=-90}
\end{center}

\figcaption[intgrl.eps]{ Ratio of general relativistic energy
deposition $\dot{Q}_{GR}$ to total newtonian energy deposition
$\dot{Q}_{Newt}$ around a neutron star as a function of neutron star
neutrinosphere radius {\bf a)} in the range relevant to supernovae
{\bf b)} and the more extreme relativistic range, down to $R = 3M$,
relevant to collapsing neutron stars. \label{intgrlratio}}

In order to show this enhancement as a function of radius, Figure
(\ref{dqdr}) plots ${d\dot{Q} \over dr}$ for several stellar masses $M
\over R$.  As expected, the enhancement is strongest near the
surface of the neutron star.  Thus the overall efficiency of this
process will be dependent upon the conditions near the surface of the
neutron star such as scale-height of the baryonic atmosphere.  This
point will be important in the discussion of supernovae.

\epsfig{file = 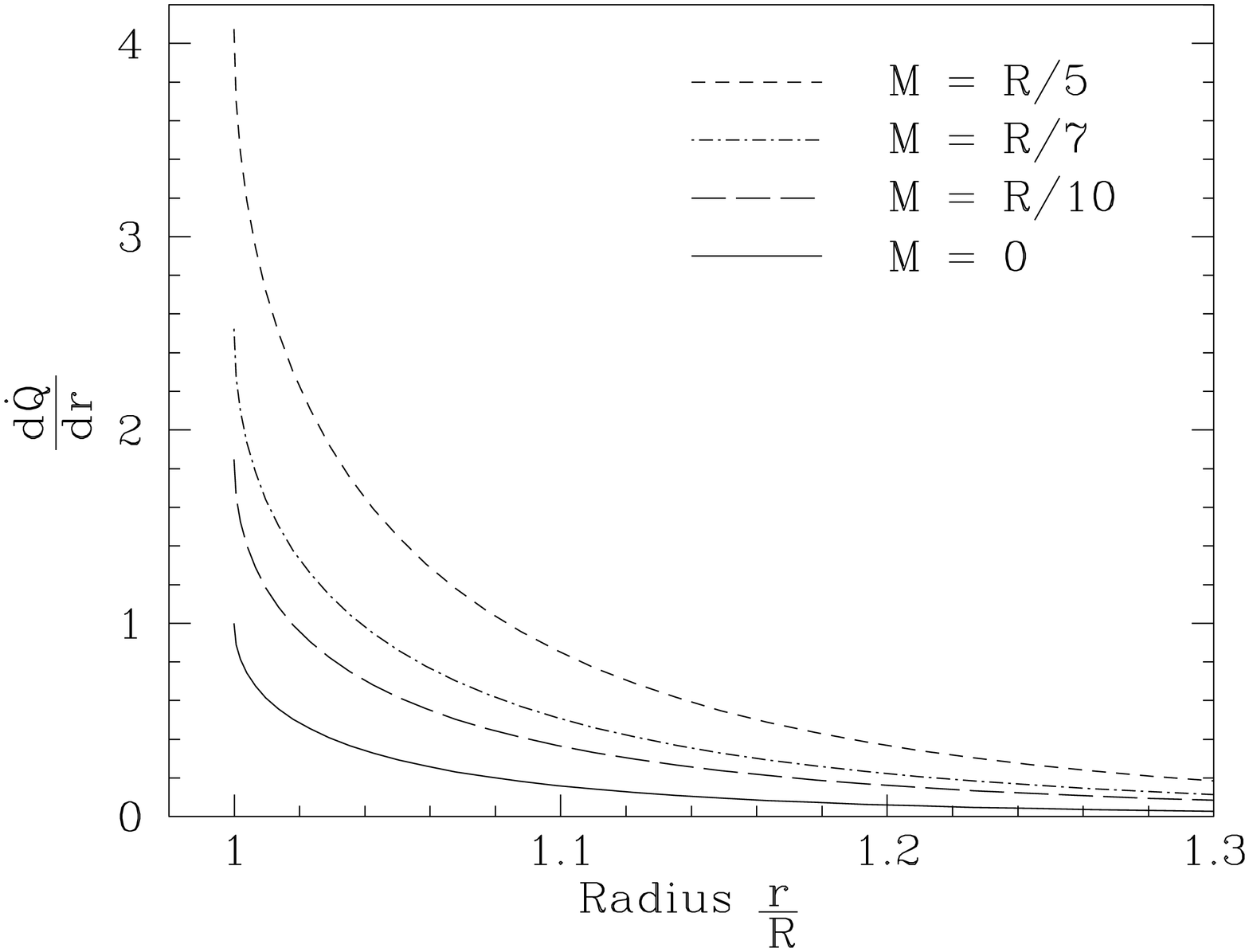, height = 15 cm, width = 7.5 cm, angle = -90}
\figcaption[dqdr.eps]{${d\dot{Q} \over dr}$ for various stellar masses
${M \over R}$.  Neutrino pair annihilation efficiency drops off
rapidly with radius.  Curves are normalized such that the Newtonian
case $M = 0$ is 1 at $r = R$.
\label{dqdr}}

The enhancements discussed here are the result of two different
relativistic effects.  The first is path bending of neutrinos
emanating from the neutron star neutrinosphere.  The
$\nu{\overline{\nu}}$ annihilation cross-section varies like the
square of the center-of-mass energy and thus favors head-on
$\nu{\overline{\nu}}$ collisions.   A simple interpretation of the
effect of path bending from Equation (\ref{E:cos}) is that it
increases the apparent angular size of the star as seen by a neutrino
at radius $r$.  This increases the number of near-horizontal neutrinos
and thus the probability of head-on collisions.

The second effect is that of gravitational redshift.  As the neutrinos
rise above the surface of the neutron star, they cool.  Since
$\nu{\overline{\nu}}$ annihilation depends strongly on energy;
$\dot{q} \propto T^9$ as in Equation (\ref{E:final1}), the
annihilation efficiency is strongly reduced by nine powers of
redshift.  However, the energy of neutrinos at the neutrinosphere of
the star is not directly observable.  Only the neutrinos at infinity
are measured.  Thus $\dot{q}$ is put in terms of an observed
luminosity $L_\infty$ in Equation (\ref{E:final2}).  The gravitational
field, then, requires a greater photosphere luminosity $L(R)$ for a
given observed luminosity $L_\infty$.  For an observed $L_\infty$
there is a net gain in $\dot{q}$ at the neutrinosphere relative to a
Newtonian calculation.  It is worthwhile to note that this effect is
important to all luminosity dependant processes near a neutron star,
not just $\nu{\overline{\nu}}$ annihilation.

\section{ Discussion }

The simple analytical arguments presented here are meant to illustrate
and quantify the importance of general relativity in the physics of
emission from compact, massive objects.  The physics near the surface
of a hot ($T \sim$ MeV) neutron star is extremely complex, with all
possible interactions between baryons, leptons, photons and neutrinos
playing important roles in energy conversion and transport.  Here we
have considered only one process: $\nu + \overline{\nu} \rightarrow
e^+ + e^-$.  This is clearly only a small component of a complete
model of the many processes involved.

However, the reaction considered here is of particular importance in
tapping the tremendous neutrino flux emanating from a hot neutron
star.  Quantifying the dependance of this reaction on gravity is
crucial to understanding the ultimate energy transport from a
radiating neutron star.

The present paper considers two problems in astrophysics for which the
efficiency of the reaction $\nu + \overline{\nu} \rightarrow e^+ +
e^-$ near the surface of a hot neutron star is important: supernovae
and close neutron star binaries.  A discussion of each in turn
follows.

\subsection{ Supernovae }

Figure (\ref{supernova}) shows the evolution of the protoneutron star
radius during a Type II supernova, starting from initial collapse, as
calculated by the Wilson-Mayle supernova code
(\cite{jay:wm88,jay:mw89}).  It is evident that the protoneutron star
shrinks rapidly for a couple of seconds. This contraction finally
halts at a radius of about 10 km.  The density scale-height at the
surface of the protoneutron star neutrinosphere is indicated in figure
(\ref{supernova}) by $|{d \log \rho \over d \log r}|$.  As the star
contracts and gravity becomes stronger, the photosphere becomes more
sharply defined.  Thus we expect the protoneutron star surface to
converge to the idealized neutrinosphere model considered here.

In order to explore this idea we define an angular cutoff factor as in
\cite{jay:cvb2}:
\begin{equation}
\chi_{Newt}(r) = {1 \over 8} (x - 1)^2 (x^2 + 4 x + 5) ~,
\label{E:chi}
\end{equation}
where $x$ is defined by Equation (\ref{E:defx}) with $M = 0$.
$\chi_{Newt}(r)$ is related to $\Theta(r)$ of Equation
(\ref{E:Thetaresult}), but only contains relative neutrino direction
information $\sim (1 - {\boldsymbol \Omega}_\nu \cdot {\boldsymbol
\Omega}_{\overline \nu})^2$ and is normalized such that $\chi = 1$ for
an isotropic neutrino distribution.  Wilson and Mayle have evaluated
the cutoff factor, $\chi_{W-M}$, in their supernova code by solving
the Boltzmann equation at various times assuming density, temperature
and electron fraction are fixed functions of radius.  Figure
(\ref{beaming}) shows a comparison between the neutrino beaming from
an idealized Newtonian photosphere and that calculated by the
Wilson-Mayle supernova code.  At early times, 0.2 seconds after the
bounce, the baryon extended atmosphere scatters the neutrinos above
the photosphere, thus rendering the neutrino angular distribution more
isotropic than is the case for an idealized sharp neutrinosphere.  At
early times, $t < 0.5$ seconds, this effect of scattering overshadows
any effect due to bending and thus simple neutrino bending is not a
valid model for these times.

\epsfig{file = 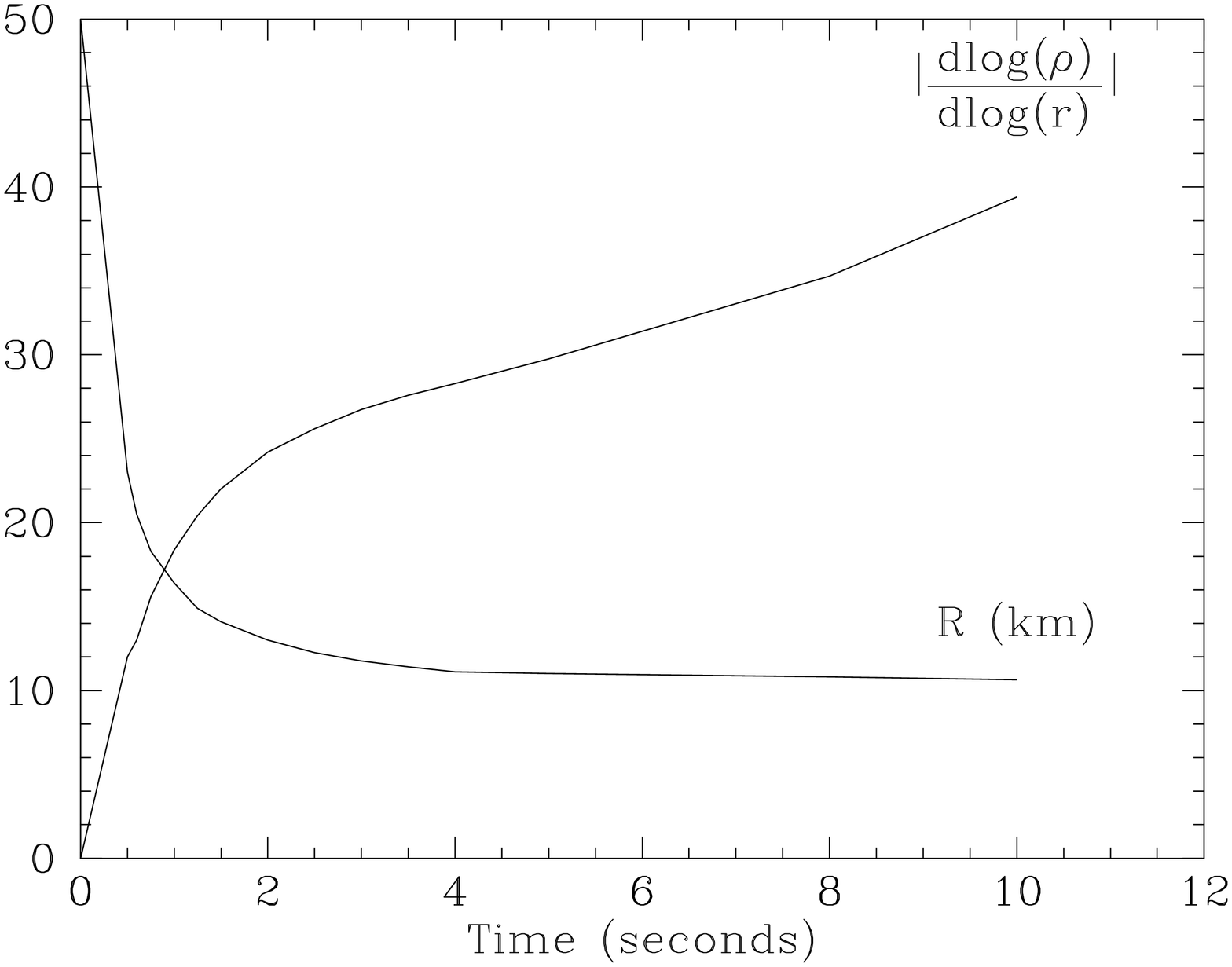, height = 15 cm, width = 7.5 cm, angle = -90}
\figcaption[ns.eps]{The evolution of a $1.41 M_\odot$ (= $2.1$ km)
protoneutron star as calculated by the the Wilson-Mayle supernova code
is characterized by rapid shrinkage to a radius $R \approx 10$ km and
a strong steepening of density profile at the photosphere as indicated
by $|{d \log \rho \over d \log r}|$. \label{supernova}}

After about one second the protoneutron star radius has contracted to
less than 20 km and its surface has become much more sharply defined,
with a baryon atmosphere scale-height $\sim 100$ meters.  Thus the
calculations exhibited here are valid for $R \lesssim 20$ km when the
star surface has converged sufficiently toward a sharp boundary and
neutrino-baryon scattering is minimized.  We conclude that after about
one second general relativistic effects are important to the dynamics
of a supernova.  In particular, at late times, $\sim 10$ seconds after
the bounce, the matter that will ultimately be involved in the
r-process is being ejected from the protoneutron star.  The heating of
the ablated matter in relativistic wind calculations (\cite{jay:cf97})
will be greatly enhanced over newtonian wind calulations
(\cite{jay:wwmhm94}).

\begin{center}
\epsfig{file = 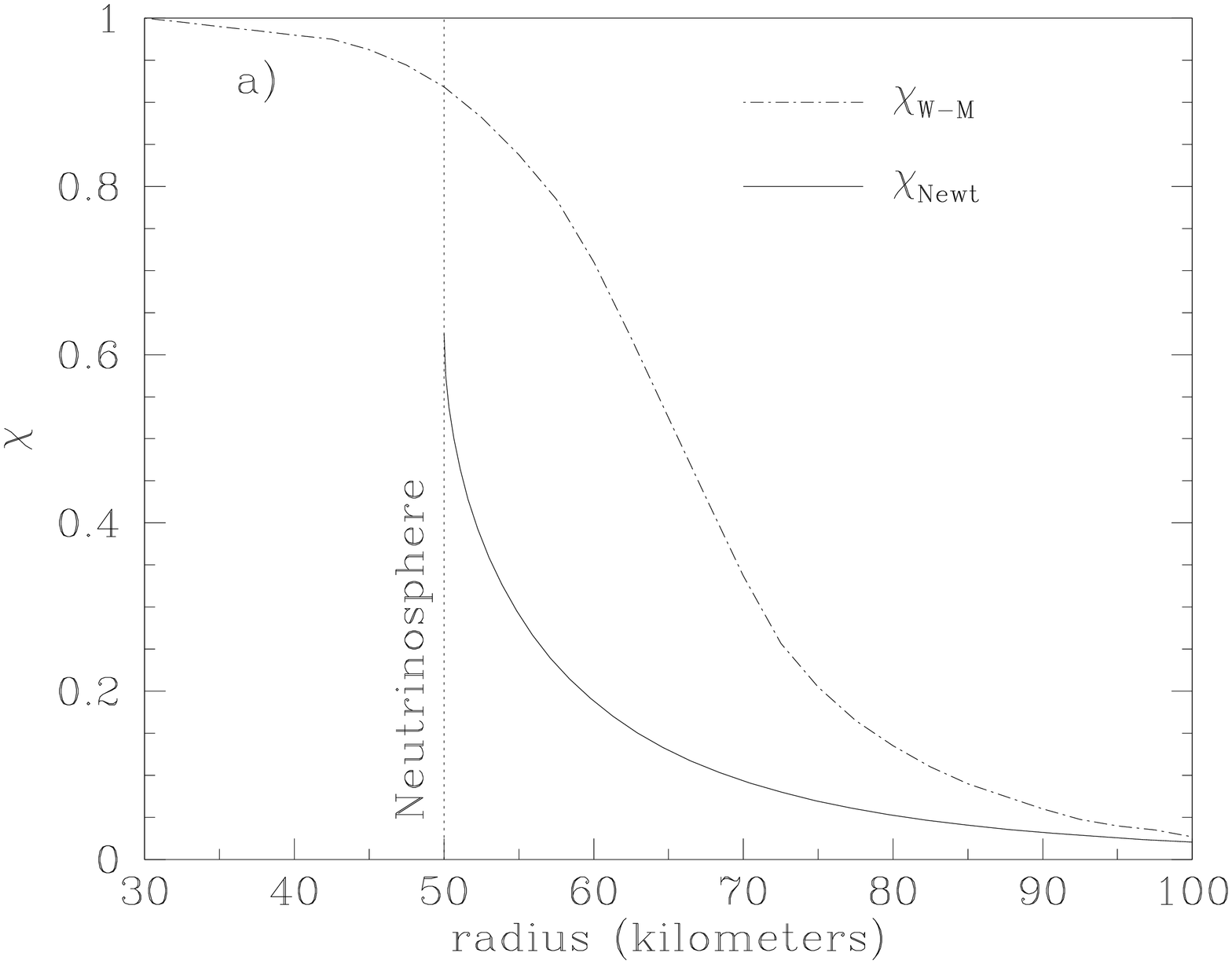, height = 15 cm, width = 7.5 cm, angle = -90}
\epsfig{file = 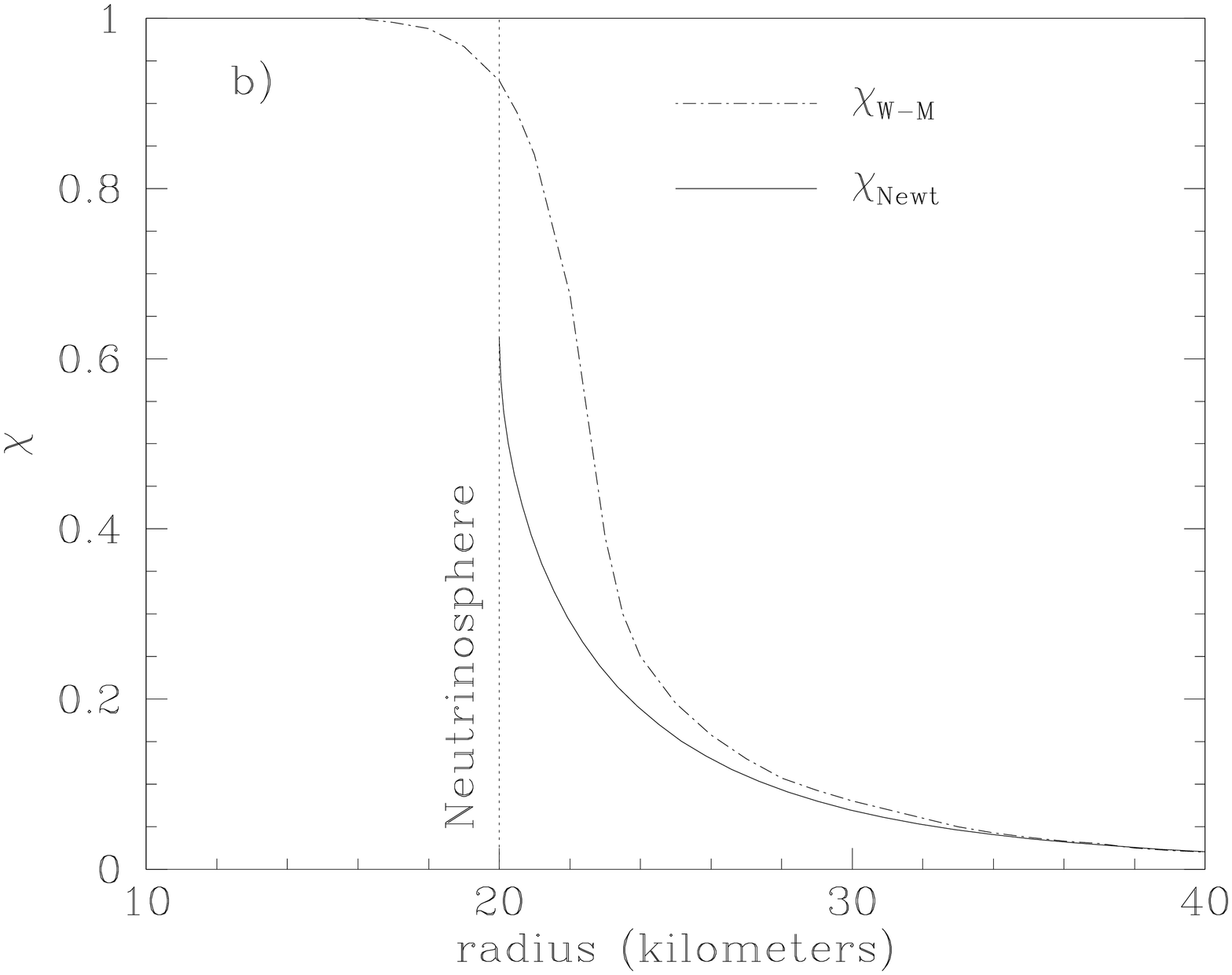, height = 15 cm, width = 7.5 cm, angle = -90}
\end{center}
\figcaption[ang50.eps]{Comparison of angular cutoff factor $\chi_{W-M}$
obtained from the Wilson-Mayle supernova code with that of the
Newtonian calculation $\chi_{Newt}$ from Equation (\ref{E:chi}).  As
the star shrinks through {\bf a)} radius $R = 50$ km at time $t = 0.2$
seconds after collapse and {\bf b)} $R = 20$ km at $t = 0.5$ seconds,
the surface becomes more sharply defined and better approximated as a
smooth neutrinosphere.
\label{beaming}}

\subsection{ Close Neutron Star Binaries }

Recent numerical calculations of neutron stars in close binary systems
near their last stable orbit have indicated that the stars may
experience a relativistic compressive force and subsequent heating
(\cite{jay:wmm96,jay:mw97,jay:mmw98}).  It is argued that $0.5 \times
10^{53}$ ergs per star of thermal neutrinos might be released within a
few seconds.  This may be a viable engine for gamma-ray bursts
(\cite{jay:wsm97}; 1998).  Accurately understanding the efficiency of
conversion from neutrinos to other forms of matter (i.e. $e^+e^-$
pairs) is crucial to predicting the possible gamma-ray burst
characteristics.  As shown in Figure (\ref{intgrlratio}), the
relativistic enhancements of the reaction $\nu\overline{\nu}
\rightarrow e^+e^-$ for a compact neutron star in a binary can be
large; up to a factor of 30 for $R = 3M$.  The calculations of
\cite{jay:wmm96} indicate that $R \approx 3M$ is the radius at which
the neutrons stars collapse to black holes.  Such large enhancements
of this reaction rate must be taken into account when modeling the
emission from a hot, compressed neutron star.  Since at late time the
neutrino luminosity rises to $\sim 10^{53}$ ergs/second, we estimate
that such enhancements will allow more than $10 \%$ of the energy
emitted in neutrinos to be converted to $e^+e^-$ pairs.  This estimate
includes electron scattering of the neutrinos by the
annihation-produced pairs, which will more than double the energy
deposition rate.  This energy in $e^+e^-$ pairs can then go on to
generate a gamma-ray burst.

\section{ Conclusions }

We have shown that the strong gravity near the surface of a neutron
star requires a relativistic treatment when modeling stellar emission.
In particular, when considering the reaction $\nu\overline{\nu}
\rightarrow e^+e^-$ the effects of path bending and redshift can
enhance the efficiency by up to a factor of 4 for $R = 5 M$ relevant
for the proto-neutron star at late-times in a type II supernovae.  A
factor of 30 is possible for $R = 3 M$ relevant to collapsing neutron
stars.  These calculations on their own are incomplete when
considering the total emission from a neutron star; one must perform
computations including all relevant processes.  However, these
calculations indicate that relativistic effects will enhance the
$e^+e^-$ pair energy from the neutrino luminosity emanating from the
neutron star.  This added $e^+e^-$ pair energy is important for
modeling late-time supernovae and for gamma-ray bursts from collapsing
neutron stars.  Thus general relativistic computations are necessary
for the accurate modeling of neutrino winds from neutron stars.

Work performed under the auspices of the U.S. Department of Energy by
the Lawrence Livermore National Laboratory under contract
W-7405-ENG-48.  

\end{document}